# Blackhole-Inspired Thermal Trapping with Graded Heat-Conduction Metadevices

Liujun Xu[1,2,3,†], Jinrong Liu[1,†], Peng Jin[1], Guoqiang Xu[2], Jiaxin Li[4], Xiaoping Ouyang[5], Ying Li[6,7], Cheng-Wei Qiu[2,*] and Jiping Huang[1,*]

[1]*Department of Physics, State Key Laboratory of Surface Physics, and Key Laboratory of Micro and Nano Photonic Structures (MOE), Fudan University, Shanghai 200438, China*

[2]*Department of Electrical and Computer Engineering, National University of Singapore, Singapore 117583, Singapore*

[3]*Graduate School of China Academy of Engineering Physics, Beijing 100193, China*

[4]*School of Mechatronics Engineering, Harbin Institute of Technology, Harbin 150001, China*

[5]*School of Materials Science and Engineering, Xiangtan University, Xiangtan 411105, China*

[6]*Interdisciplinary Center for Quantum Information, State Key Laboratory of Modern Optical Instrumentation, ZJU-Hangzhou Global Scientific and Technological Innovation Center, Zhejiang University, Hangzhou 310027, China*

[7]*International Joint Innovation Center, Key Laboratory of Advanced Micro/Nano Electronic Devices and Smart Systems of Zhejiang, The Electromagnetics Academy of Zhejiang University, Zhejiang University, Haining 314400, China*

[†]*These authors contributed equally to this work.*

[*]*Corresponding authors. E-mails: chengwei.qiu@nus.edu.sg; jphuang@fudan.edu.cn*

# Abstract

Black holes are one of the most intriguing predictions of general relativity. So far, metadevices have enabled analogous black holes to trap light or sound in laboratory spacetime. However, trapping heat in a conductive ambient is still challenging because diffusive behaviors are directionless. Inspired by black holes, we construct graded heat-conduction metadevices to achieve thermal trapping, resorting to the imitated advection produced by graded thermal conductivities rather than the trivial solution of using insulation materials to confine thermal diffusion. We experimentally demonstrate thermal trapping for guiding hot spots to diffuse towards the center. Graded heat-conduction metadevices have advantages in energy-efficient thermal regulation because the imitated advection has a similar temperature field effect to the realistic advection that is usually driven by external energy sources. These results also provide insights into correlating transformation thermotics with other disciplines such as cosmology for emerging heat control schemes.

# Introduction

Efficient heat utilization is crucial for industrial production and daily life [1-3]. However, heat control is still facing many challenges because thermal diffusion, a primary heat transfer scheme, is directionless. Thermal advection, another fundamental heat transfer mode, can break the intrinsic space-reversal symmetry of thermal diffusion and realize asymmetric heat transfer. However, both the actual advection induced by mass transfer [4-6] and the effective advection generated by spatiotemporal modulation [7-10] require the implementation of inconvenient and energy-consuming external drives. Therefore, it is not only a different subject but also a big challenge to realize asymmetric diffusion in a pure and passive heat-conduction environment.

On the other side, black holes are well-known for their ability to trap light in their horizons. Due to the development of metadevices, analogous black holes have been explored in electromagnetics [11-17] and acoustics [18-21], thus trapping light and sound in laboratory spacetime. However, because of the vast difference between wave and diffusion, these analogous black holes have no natural correspondence in heat conduction. Therefore, it is still intrinsically elusive to realize thermal trapping.

Inspired by black holes, we propose purely heat-conduction-based metadevices with graded thermal parameters [22-24] to realize thermal trapping. Though the designed structures are stationary solids without external drives, the graded thermal conductivities can still generate the imitated advection that brings about a similar temperature field effect to the realistic advection induced by mass transfer. As an alternative understanding, graded parameters can also be mathematically linked to curvilinear spacetime by the transformation theory [25-28]. Then we design the imitated advection to point towards the center, and surrounding heat can also be trapped towards the center, thereby achieving thermal trapping. Inspired by rotational black holes, we further perform a rotation transformation to realize rotational thermal trapping. These schemes are validated numerically and experimentally.

## The imitated advection

The space-reversal symmetry makes thermal diffusion directionless. Specifically, hot spots in a homogeneous structure only dissipate but do not move (Fig. 1a and b). An insulator-conductor-insulator structure can introduce anisotropy [29] and make thermal diffusion distinctly different in two perpendicular directions: one direction very conductive and the other very insulating. Nevertheless, if a hot spot is in the conductive region bounded by insulation walls, heat still goes forward and backward along the conductive pathway (the right-top and left-bottom hot spots in Fig. 1c and d). If a hot spot is in the region with a very low thermal conductivity, heat diffuses slightly in an isotropic way (the left-top and right-bottom hot spots in Fig. 1c and d).

In contrast to these conventional structures, a graded structure can break the space-reversal symmetry of thermal diffusion and trap hot spots towards the center even with rotation (Fig. 1e and f). The underlying mechanism lies in the imitated advection induced by graded thermal conductivities. We compare wave propagation and heat diffusion to understand the imitated advection. In photonics, a graded refractive index can generate an effective momentum [30-32]. Hence, the bending effect can be achieved despite the vertical incidence of waves (Fig. 2a). The imitated advection (Fig. 2b) shares a similar conception with the effective momentum.

Heat conduction in one dimension is described by $\rho_0 C_0 \partial_t T + \partial_x(-\kappa_0 \partial_x T) = 0$, where $\rho_0$, $C_0$, and $\kappa_0$ are the mass density, heat capacity, and thermal conductivity of a homogeneous medium. $T$, $t$, and $x$ denote temperature, time, and coordinate. To achieve the imitated advection, we consider a graded thermal conductivity of $\kappa(x)$, a graded mass density of $\rho(x)$, and a graded heat capacity of $C(x)$. Hence, the heat conduction equation becomes $\rho(x)C(x)\partial_t T + \partial_x(-\kappa(x)\partial_x T) = 0$, which is further reduced to

$$\frac{\partial T}{\partial t} - \frac{1}{\rho(x)C(x)} \frac{\partial \kappa(x)}{\partial x} \frac{\partial T}{\partial x} - \frac{\kappa(x)}{\rho(x)C(x)} \frac{\partial^2 T}{\partial x^2} = 0. \tag{1}$$

We consider the diffusion-advection equation of $\partial_t T + v_0 \partial_x T - D_0 \partial_x^2 T = 0$, where $v_0$ is the advection velocity, and $D_0 = \kappa_0/(\rho_0 C_0)$ is the thermal diffusivity of a homogeneous medium. Hence, the second term on the left side of Eq. (1) denotes the imitated advection. The imitated advection velocity of $v_i$ is

$$v_i = \frac{-1}{\rho(x)C(x)} \frac{\partial \kappa(x)}{\partial x}. \tag{2}$$

The graded thermal conductivity with $\partial_x \kappa(x) \neq 0$ is the key to the imitated advection. For a constant velocity, we assume the parameters have exponential forms,

$$\kappa(x) = \kappa_0 e^{\alpha x}, \tag{3a}$$

$$\rho(x)C(x) = \rho_0 C_0 e^{\alpha x}, \tag{3b}$$

where $\alpha$ is a constant. The substitution of Eqs. (3a) and (3b) into Eq. (2) yields a constant velocity of $v_{i0}$,

$$v_{i0} = -\alpha D_0. \tag{4}$$

On the one hand, Eqs. (3a) and (3b) lead to a global constant thermal diffusivity of $D_0 = \kappa(x)/(\rho(x)C(x)) = \kappa_0/(\rho_0 C_0)$. Hence, the designed structure is still homogeneous on a thermal-diffusivity level. In general, thermal diffusivity reflects the ability of heat to diffuse, and a constant value means the same diffusive properties, thereby making the imitated advection unexpected. On the other hand, Eqs. (3a) and (3b) ensure a constant imitated advection velocity. Since the imitated advection is attributed to graded thermal conductivities, the exponential variation is not mandatory as long as $\partial_x \kappa(x) \neq 0$.

The imitated advection has nearly the same temperature field effect as the realistic advection (Supplementary Part I), so it can also break the space-reversal symmetry of thermal diffusion and generate asymmetric temperature profiles. This equivalence is only on a temperature-field level rather than a heat-flux level because graded thermal conductivities cannot replace fluid flows in fluid dynamics.

To demonstrate the imitated advection, we fabricate a graded heat-conduction metadevice, i.e., a copper plate drilled with air holes (Fig. 2c). The effective graded thermal conductivity of the fabricated sample $\kappa_e$ is determined by the Maxwell-Garnett formula $\kappa_e = (\kappa_1 + \kappa_2 + (\kappa_1 - \kappa_2)f)/(\kappa_1 + \kappa_2 - (\kappa_1 - \kappa_2)f)$, where $\kappa_1$ and $\kappa_2$ are the thermal conductivities of air and copper, and $f$ is the area fraction of the air holes [33]. Hence, we can use only two natural materials to achieve a graded thermal conductivity with graded hole size. The theoretical distribution of the graded thermal conductivity is $\kappa_e/\kappa_0 = e^{-1.4x/L}$, where $L$ is the structure length (Fig. 2d).

We define the normalized temperature as $T^* = T/(T_h - T_c) - T_c/(T_h - T_c)$, where $T$ is the realistic temperature, and $T_h$ (or $T_c$) is the temperature of the hot (or cold) source. For thermal diffusion in a homogeneous medium, the forward and backward cases are symmetric, with temperature distributions of $T^* = -x/L + 1$ and $T^* = x/L$, respectively. Therefore, $T^* = 0.5$ should appear at $x/L = 0.5$. Intriguingly, the graded heat-conduction metadevice can produce the imitated advection and break the space-reversal symmetry of heat conduction. Hence, the measured temperature profiles (Fig. 2e and f) demonstrate that $T^* = 0.5$ is off-center. For clarity, we plot the temperature distributions in Fig. 2g. For the forward and backward cases, $T^* = 0.5$ appears at the positions of $x/L = 0.6$ and $x/L = 0.73$, respectively. The solid lines represent the theoretical results with ideal graded parameters. The dashed and dotted lines demonstrate the results based on the finite-element method (FEM) and experimental measurement (Exp.). The theory, simulations, and experiments are almost in accordance. We further show the simulated temperature profiles with the realistic advection (Fig. 2h and i), and $T^* = 0.5$ appears at the same positions of $x/L = 0.66$ for forward and backward cases. Therefore, the temperature field effect of the imitated advection is indeed similar to that of the realistic advection.

Then we explain why $T^* = 0.5$ does not appear at the same positions in Fig. 2e and f (theoretically the same). The measured forward and backward temperature curves are always lower than the simulated temperature curves (Fig. 2g) because there is energy loss resulting from thermal radiation and natural convection in experiments. The energy loss shifts the position with $T^* = 0.5$ to the left for the forward case and to the right for the backward case, agreeing with the experimental results in Fig. 2e and f. More importantly, the energy loss makes the forward and backward heat fluxes asymmetric because the temperature gradients are no longer identical at the same positions in Fig. 2e and f. This energy-loss-induced heat-flux asymmetry indicates potential correspondence to the loss-induced nonreciprocal wave propagation [34,35].

Therefore, we can draw two conclusions from Fig. 2. One is that we experimentally demonstrate the imitated advection in a pure and passive heat-conduction environment, which is similar to the realistic advection in terms of the temperature field effect. The other is that energy loss in graded heat-conduction metadevices can further generate asymmetric heat fluxes.

## Blackhole-inspired thermal trapping

Inspired by black holes, we guide the imitated advection to point towards the center to realize thermal trapping (Supplementary Part II). The one-dimensional conclusions also apply to two and three dimensions (Supplementary Part III). Therefore, we are allowed to design two-dimensional thermal trapping in what follows. We also provide a qualitative explanation of the physical basis for linking graded parameters and curvilinear spacetime with the conformal transformation theory (Supplementary Part IV).

We first perform two-dimensional simulations to show directionless thermal diffusion in a homogeneous medium by COMSOL Multiphysics (Fig. 3a-c). The four hot spots are almost stationary, indicating no trapping effect in a homogeneous medium. Then we design the imitated advection (represented by the dashed arrows in Fig. 3d) to be centrally pointing, and the parameters are radially distributed, e.g., the thermal conductivity of $\kappa(r)/\kappa_0 = e^{\alpha(1-r/R)}$, where $r$ denotes the radial coordinate, and $R$ is the radius of the circular plate. Hence, the four hot spots can be trapped towards the center (Fig. 3e and f).

Inspired by rotational black holes, we perform a rotation transformation on the scheme shown in Fig. 3d to demonstrate rotational thermal trapping (Fig. 3g). The rotation transformation is $r = r'$ and $\theta = \theta' + b(r - R)$, where $(r, \theta)$ and $(r', \theta')$ are cylindrical coordinates in physical and virtual spaces, respectively. We define $b = \theta_0(R_0 - R)$, where $\theta_0$ is the rotation angle, and $R_0$ is a constant radius. With the transformation thermotics theory [36-38], the transformed parameters are $\overleftrightarrow{\kappa} = J\kappa'J^\dagger/\det J$ and $\rho C = \rho'C'/\det J$, where the superscript of ′ denotes those parameters in virtual space, and $J^\dagger$ and $\det J$ are the transpose and determinant of $J$, respectively. Since $J$ is the Jacobian transformation matrix expressed as $J = [(\partial r/\partial r', \partial r/(r'\partial\theta')), (r\partial\theta/\partial r', r\partial\theta/(r'\partial\theta'))]$, the transformed thermal conductivity for rotational thermal trapping is a tensor, not only graded but also anisotropic. The effect of rotational thermal trapping can be observed in Fig. 3h and i.

The simulated temperature evolution of these two types of thermal trapping is also exported as animations (Supplementary Gifs I for normal thermal trapping and II for rotational thermal trapping). Besides, thermal trapping can still be achieved with periodic temperature excitation (Supplementary Gifs III for normal thermal trapping and IV for rotational thermal trapping) because the major mechanism is the imitated advection generated by graded parameters, which is robust against different excitations.

We fabricate two samples (Fig. 4a and d) to demonstrate thermal trapping. Three common materials (i.e., copper, iron, and steel) are used to increase the thermal conductivity gradient to $\sim e^{-4.8(1-r/R)}$. We use four bigger initial hot spots to reduce the effect of heat dissipation (Fig. 4a and d). For normal thermal trapping, the hot spots are trapped towards the center directly (Fig. 4b and c). For rotational thermal trapping, the hot spots are both trapped and rotated (Fig. 4e and f). To measure these two samples, we put them into an ice water tank (acting as a cold source, Fig. 4g). Heat guns generate hot spots, which can output a constant temperature for the initial setting. Two types of thermal trapping can be observed in Fig. 4h and i and Fig. 4j-l. Two factors mainly affect the experimental results. The first factor is the heat dissipation by natural convection and thermal radiation, making these hot spots decay too fast to observe. The second factor is the approximate parameters restricted by materials. Since our proof-of-concept experiments realize a thermal conductivity gradient much lower than the simulations in Fig. 3, the trapping effect is not as ideal as the simulations in Fig. 3. Nevertheless, these two factors do not affect our main conclusions.

## Discussion and conclusion

Thermal advection is crucial for non-Hermitian physics and nonreciprocal transport, helping reveal various intriguing thermal phenomena such as exceptional points [39], topological transitions [40], and diffusive Fizeau drag [41]. Since the imitated advection has almost the same temperature field effect as the realistic advection, it is promising to uncover similar phenomena [39-41] with graded heat-conduction metadevices. The major advantage is zero energy consumption because external drives are not required. Moreover, due to the asymmetric thermal diffusion induced by the imitated advection, graded heat-conduction metadevices have potential applications in thermal funneling [42]. The imitated advection may also provide insights into controlling near-field thermal radiation [43-45] and fluid flows [46].

To summarize, we have revealed blackhole-inspired thermal trapping with graded heat-conduction metadevices. The underlying mechanism lies in the imitated advection induced by graded thermal conductivities as a counterpart to graded refractive indexes responsible for the effective momentum in photonics. In terms of the temperature field effect, the imitated advection can replace the realistic advection induced by mass transfer and yield asymmetric temperature profiles. Moreover, energy loss in experiments can facilitate asymmetric heat fluxes. Finite-element simulations and laboratory experiments have confirmed the performance of thermal trapping, contributing to waste heat recovery. Our work brings new thoughts and perspectives linking diffusive systems (e.g., thermotics, particle dynamics, etc.), wave systems (e.g., photonics, acoustics, etc.), and cosmological phenomena [47-50].

# Methods

## Simulation

All finite-element simulations are performed with the commercial software COMSOL Multiphysics. The simulations in Figs. 1-4 are based on the template of heat transfer in solids. When we compare the imitated advection with the realistic advection in Fig. 2 and Fig. S1, the template of heat transfer in fluids is also applied.

## Fabrication

The samples are prepared with a milling machine. Three materials are finely joined together, reducing the effect of thermal interfacial resistance as much as possible. The samples are also covered with plastic films to reduce thermal radiation reflection. We use an infrared camera to detect the temperature profiles. The hot and cold sources in Fig. 2 are water baths. Heat guns generate the initial hot spots in Fig. 4.

## Supplementary data

Supplementary data are available at NSR online.

## Funding

This work was supported by the National Natural Science Foundation of China to J.H. (11725521 and 12035004), the Science and Technology Commission of Shanghai Municipality to J.H. (20JC1414700), the Singapore Ministry of Education to C.-W.Q. (R-263-000-E19-114), and the National Natural Science Foundation of China to Y.L. (92163123).

## Author contributions

L.X., C.-W.Q., and J.H. conceived of the idea. L.X. proposed the methodology, performed the derivations, and designed the experiments. J.L. and P.J. conducted the experiments. L.X., J.L.,

P.J., G.X., J.L., X.O., Y.L., C.-W.Q., and J.H. made the visualization and wrote the manuscript. C.-W.Q. and J.H. supervised the work. All authors contributed to the discussion and finalization of the manuscript.

***Conflict of interest statement.*** None declared.

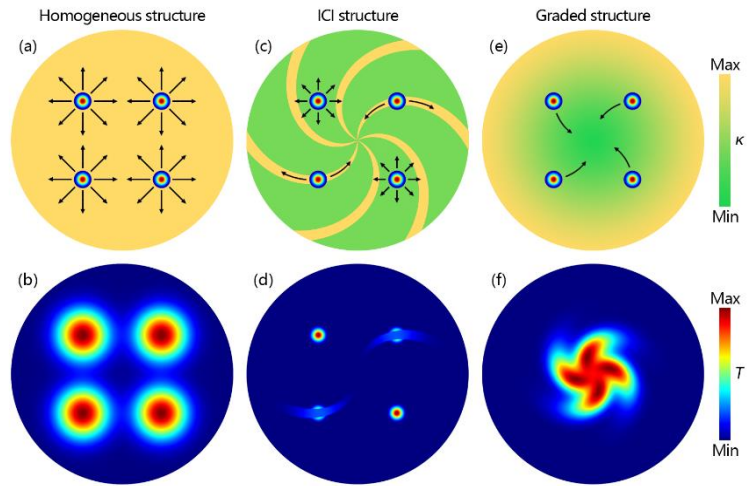

**Figure 1.** Thermal diffusion in different structures. (a) Homogeneous structure with isotropic diffusion. (b) Temperature distribution of (a) at a specific moment. (c) and (d) Insulator-conductor-insulator (ICI) structure with anisotropic diffusion in yellow strips (high thermal conductivity) and isotropic diffusion in green areas (low thermal conductivity). (e) and (f) Graded structure with asymmetric diffusion towards the center.

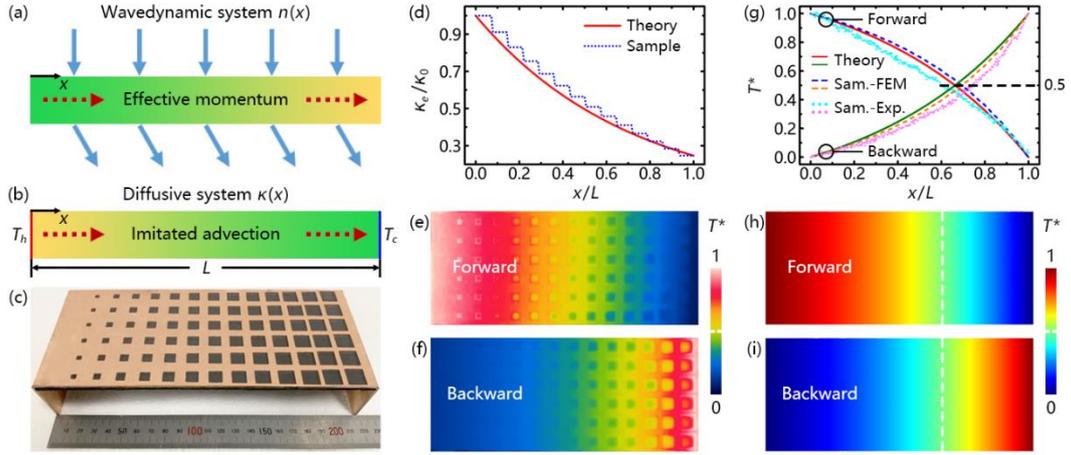

**Figure 2.** Experimental demonstration of the imitated advection. (a) Effective momentum generated by a graded refractive index of $n(x)$. (b) Imitated advection induced by a graded thermal conductivity of $\kappa(x)$. (c) Photo of the fabricated graded heat-conduction metadevice. (d) Graded thermal conductivity distributions. Measured temperature profiles with (e) forward and (f) backward thermal fields. (g) Normalized temperature distributions. Simulated temperature profiles with the actual advection in the presence of (h) forward and (i) backward thermal fields. The white dashed lines represent the positions with $T^* = 0.5$. The thermal conductivity, mass density, and heat capacity of copper or air are $\kappa_{copper} = 400$ W m$^{-1}$ K$^{-1}$, $\rho_{copper} = 8900$ kg/m$^3$, and $C_{copper} = 390$ J kg$^{-1}$ K$^{-1}$; $\kappa_{air} = 0.026$ W m$^{-1}$ K$^{-1}$, $\rho_{air} = 1.29$ kg/m$^3$, and $C_{air} = 1000$ J kg$^{-1}$ K$^{-1}$, respectively. Sam.: Sample; FEM: Finite-Element Method; Exp.: Experiment.

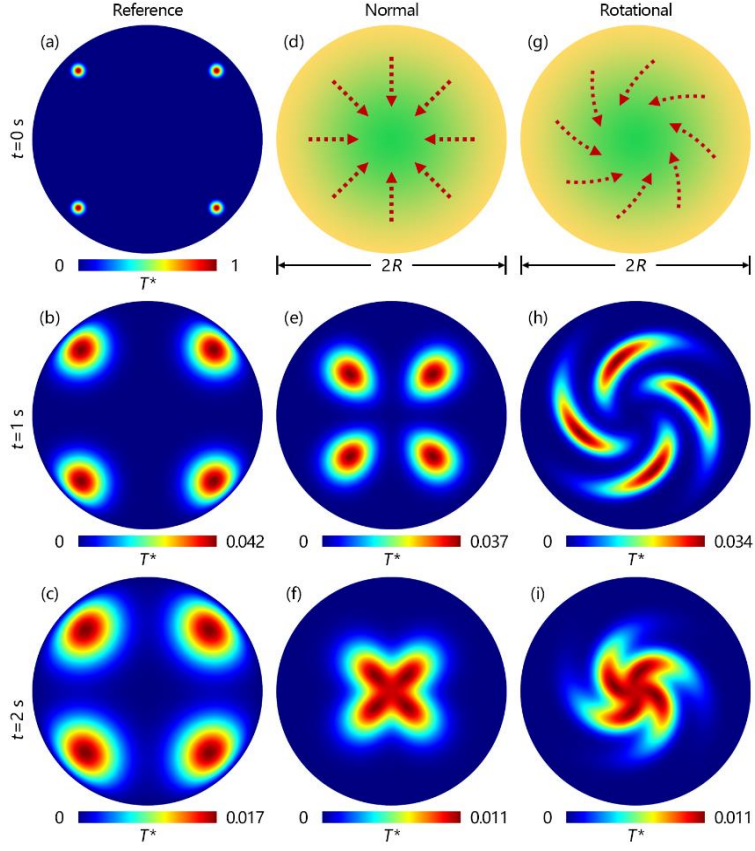

**Figure 3.** Simulations of thermal trapping. (a)-(c) Temperature evolution in a homogeneous plate. The right-top hot spot is at $(a, a)$ with $T^* = e^{-\left(200\sqrt{(x-a)^2+(y-a)^2}\right)^2}$ and $a = 0.06$ m. The outer boundaries are set at $T^* = 0$. (d)-(f) Normal thermal trapping with parameters of $\kappa(r)/\kappa_0 = e^{-30(1-r/R)}$, $\rho(r)/\rho_0 = e^{-30(1-r/R)}$, $\kappa_0 = 400$ W m$^{-1}$ K$^{-1}$, $\rho_0 = 8900$ kg/m$^3$, $C_0 = 390$ J kg$^{-1}$ K$^{-1}$, and $R = 0.1$ m. (g)-(i) Rotational thermal trapping with parameters of $\overleftrightarrow{\kappa}(r)/\kappa_0 = [(1, br), (br, b^2r^2 + 1)]e^{-30(1-r/R)}$, $\rho(r)/\rho_0 = e^{-30(1-r/R)}$, and $b = -32$ m$^{-1}$.

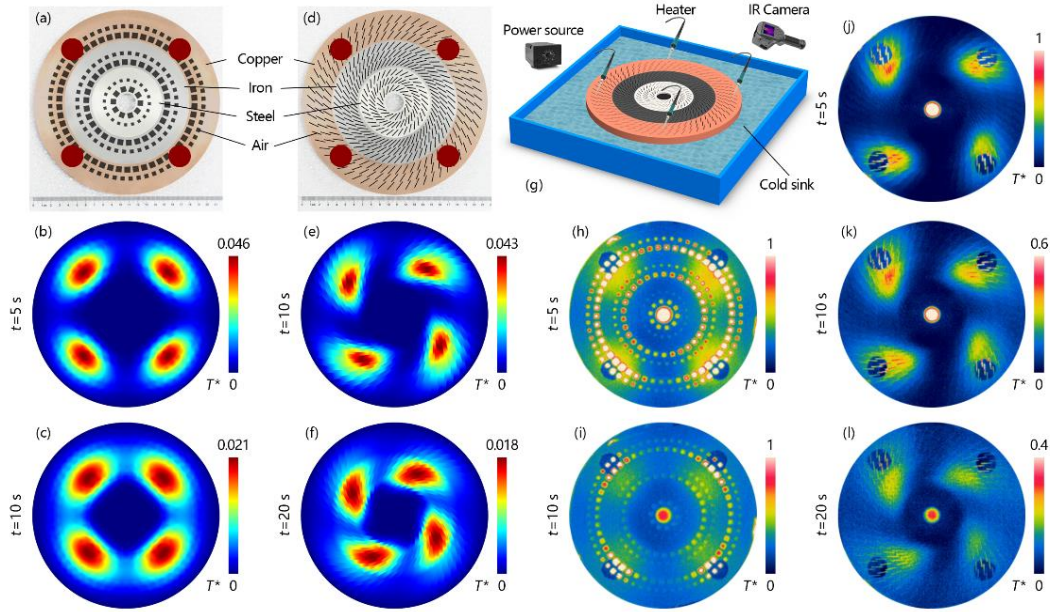

**Figure 4.** Experiments of thermal trapping. (a) and (d) Two sample photos. Simulations of (a) at (b) 5 s and (c) 10 s. Simulations of (d) at (e) 10 s and (f) 20 s. The initial four hot spots with normalized temperatures of $T^* = 1$ and radii of 12 mm are also presented in (a) and (d). (g) Experimental setup schematic. Measured temperature profiles of (h) and (i) normal and (j)-(l) rotational thermal trapping. The thermal conductivity, mass density, and heat capacity of iron or steel are $\kappa_{iron} = 80$ W m$^{-1}$ K$^{-1}$, $\rho_{iron} = 7800$ kg/m$^3$, and $C_{iron} = 460$ J kg$^{-1}$ K$^{-1}$; $\kappa_{steel} = 15$ W m$^{-1}$ K$^{-1}$, $\rho_{steel} = 8000$ kg/m$^3$, and $C_{steel} = 500$ J kg$^{-1}$ K$^{-1}$, respectively.